\documentclass[a4paper,12pt]{article}
\usepackage[T1]{fontenc}
\usepackage[ragged]{footmisc}
\usepackage[left=1in,right=1in,top=1in,bottom=1in]{geometry}
\usepackage{color}
\usepackage{tabularx}
\usepackage{multirow}

\usepackage{adjustbox}

\makeatletter
\renewcommand\footnotesize{%
   \@setfontsize\footnotesize\@ixpt{11}%
   \abovedisplayskip 8\p@ \@plus2\p@ \@minus4\p@
   \abovedisplayshortskip \z@ \@plus\p@
   \belowdisplayshortskip 4\p@ \@plus2\p@ \@minus2\p@
   \def\@listi{\leftmargin\leftmargini
               \topsep 4\p@ \@plus2\p@ \@minus2\p@
               \parsep 2\p@ \@plus\p@ \@minus\p@
               \itemsep \parsep}%
   \belowdisplayskip \abovedisplayskip
}
\makeatother

\usepackage{epigraph} 
\usepackage{setspace}  
\usepackage{comment}

\usepackage[english,activeacute]{babel} 
\usepackage[T1]{fontenc}				
\usepackage[utf8]{inputenc}				
\usepackage{tgpagella}					

\usepackage{etoolbox}
\usepackage{appendix}
\AtBeginEnvironment{appendices}{\crefalias{section}{appendix}}

\usepackage{tabularx}
\usepackage[flushleft]{threeparttable}
\usepackage{longtable}
\usepackage{booktabs}
\usepackage{graphicx}
\usepackage{marvosym}
\usepackage[super]{nth}
\usepackage{fancyvrb}  
\usepackage{epstopdf}
\newcommand{\sym}[1]{#1} 
\usepackage{pdflscape}
\usepackage{pdflscape}
\usepackage{subcaption}
\usepackage[justification=centering]{caption}
\usepackage{float}

\usepackage{multicol}
\usepackage{amsmath,amsfonts,amssymb}	
\usepackage{dsfont}						
\usepackage{newpxtext} 
\usepackage{newpxmath} 

\usepackage{breqn}

\usepackage{dsfont}
\usepackage[dvipsnames]{xcolor}
\definecolor{darkblue}{rgb}{0,0,102}
\usepackage[unicode=true,pdfusetitle,bookmarks=true,bookmarksnumbered=false,bookmarksopen=false,breaklinks=false,pdfborder={0 0 0},backref=false,colorlinks=true,linkcolor=darkblue,citecolor=darkblue]{hyperref}

\usepackage{geometry}
\usepackage{longtable} 
\usepackage{adjustbox}

\usepackage{natbib}                 
\usepackage[nameinlink,capitalize,noabbrev]{cleveref}
\usepackage{soul} 

\usepackage[normalem]{ulem}
\useunder{\uline}{\ul}{}

\usepackage{titlesec}
\usepackage{hyperref}
\hypersetup{
    colorlinks=true,
    linkcolor=blue,
    filecolor=magenta,      
    urlcolor=cyan,
    pdftitle={Overleaf Example},
    pdfpagemode=FullScreen,
    }



\begin{document}


\title{\textbf{The Aftermath of Peace: The Impact of the FARC's Ceasefire on Forced Displacement in Colombia}\thanks{We thank Marianna Battaglia, Lola Collado, Marinella Leone, and Jaime Millán-Quijano for their comments. Albarrán and Sanz-de-Galdeano acknowledge financial support from Project PID2021-124237NB-I00 (financed by MCIN/ AEI /10.13039/501100011033/ and by FEDER Una manera de hacer Europa) and from Generalitat Valenciana, Consellería de Innovación, Universidades, Ciencia y Sociedad Digital through project Prometeo CIPROM/2021/068.} \\}

\author{Pedro Albarrán \thanks{Corresponding author. Department of Economics (FAE), University of Alicante. {albarran@ua.es}.} \and Antonio Robles\thanks{Department of Economics (FAE), University of Alicante. {a.roblesr85@gmail.com}.} \and Anna Sanz-de-Galdeano\thanks{Department of Economics (FAE), University of Alicante and IZA. {anna.sanzdegaldeano@gmail.com}.}
}

\date{\today }

\maketitle

\begin{singlespace}
\begin{abstract}
Colombia’s prolonged conflict has made the country one of the most affected by forced internal displacement (FID) in the world. This study examines the impact of the FARC’s 2014 unilateral and permanent ceasefire on FID. We use a difference-in-differences strategy that exploits the timing of the ceasefire and the pre-conflict distribution of FARC presence across municipalities. Results show a substantial reduction in severe displacement episodes in affected areas, with effects that emerged gradually and persisted over time. These findings highlight the importance of stability and the effective implementation of peace agreements in mitigating FID and its far-reaching consequences.

\end{abstract}
\end{singlespace}
\newpage

\section{Introduction} \label{Sec:intro}

According to the United Nations High Commissioner for Refugees (UNHCR) \textit{Global Trends} report, “At the end of 2023, an estimated 117.3 million people worldwide were forcibly displaced due to persecution, conflict, violence, human rights violations and events seriously disturbing the public order.” Forced migration has wide-ranging effects on receiving populations, displaced individuals themselves, and sending communities \citep{becker2019consequences, verme2021impact}. One form of forced displacement occurs when people “have been forced to flee their homes by conflict, violence, persecution or disasters, however, they remain within the borders of their own country”—a phenomenon known as forced internal displacement (FID), as defined by the UNHCR. By June 2024, 72.1 million people were internally displaced, accounting for the majority (59\%) of the world’s forcibly displaced population.\footnote{\url{https://www.unhcr.org/about-unhcr/who-we-protect/internally-displaced-people}.}

The Internal Displacement Monitoring Centre (IDMC) reports that “\textit{Colombia has faced one of the world’s most acute internal displacement situations associated with conflict and violence for five decades.}” According to IDMC’s 2020 Global Report on Internal Displacement, Colombia had the second-largest number of internally displaced persons at the end of 2019 (5.5 million) just behind Syria \citep{grid2020}. This large-scale displacement is primarily rooted in the country’s prolonged internal conflict. The emergence of left-wing guerrilla groups in the 1960s, such as the \textit{Fuerzas Armadas Revolucionarias de Colombia} (FARC, Revolutionary Armed Forces of Colombia) and the \textit{Ejército de Liberación Nacional} (ELN, National Liberation Army), led to escalating violence. In the 1980s, drug cartels (notably Medellín and Cali) and right-wing paramilitary groups further intensified the conflict, displacing millions to seize land for agro-industry, mining, and drug trafficking.

\begin{figure}[H]
    \centering
        \caption{Forced Internal Displacement in Colombia 1994-2019}
    \label{fig:fdi_col}
    \includegraphics[width=0.75\linewidth]{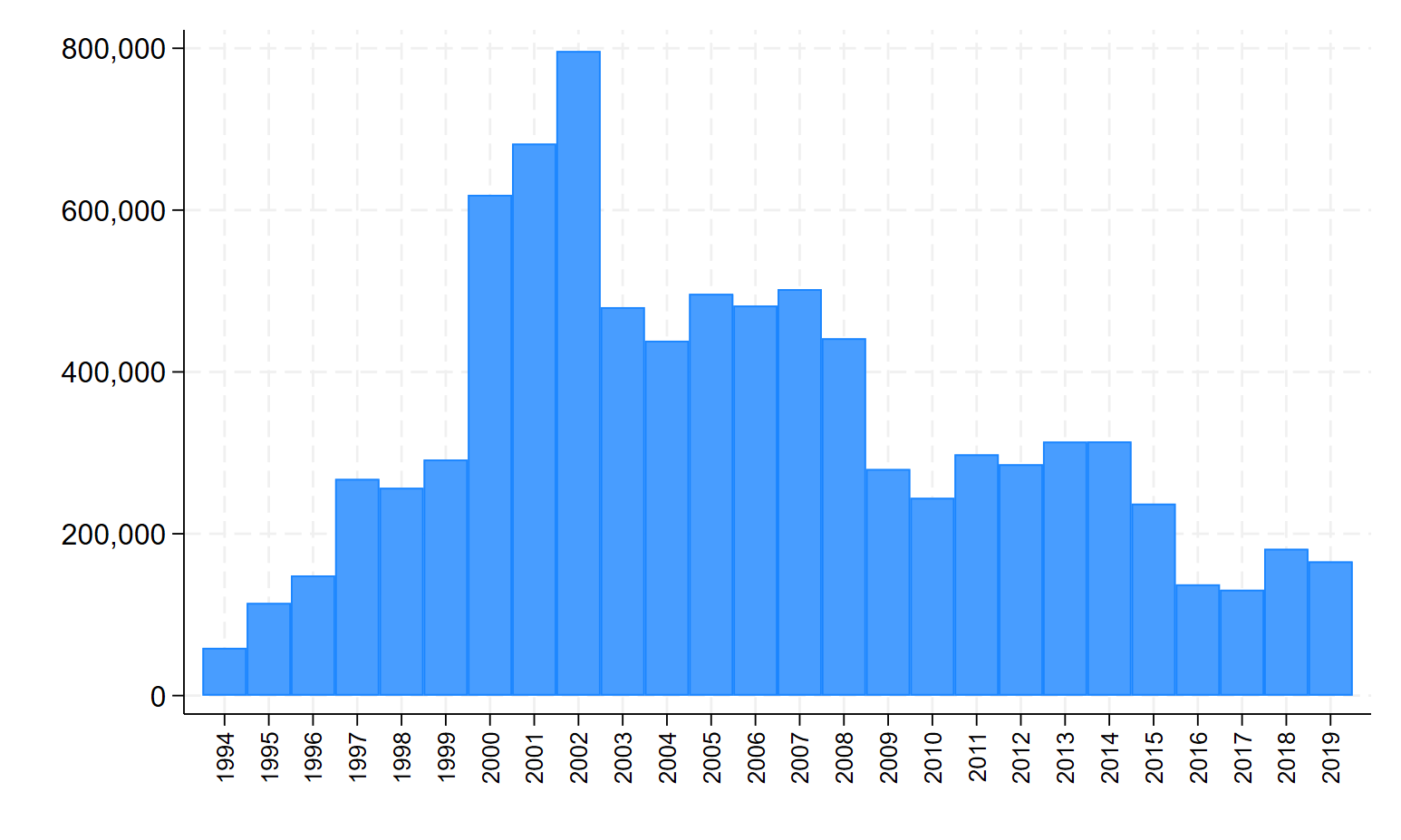}
    \begin{minipage}{0.75\linewidth} 
\centering \footnotesize Source: Author's calculations using data from the \textit{Registro Único de Víctimas}.
\end{minipage}\\
\end{figure}

\autoref{fig:fdi_col} shows the total number of people forcibly internally displaced in Colombia each year from 1994 to 2019. Displacement peaked at nearly 800,000 individuals in 2002, during the most violent period.

The patterns of FID in Colombia have had profound and lasting consequences. Displacement entails the loss of one’s way of life, including jobs, property, and vital social networks. Empirical studies have documented a wide range of negative outcomes: \citet{ibanez2008civil} estimate a 37\% decline in the net present value of lifetime rural consumption; \citet{wharton2011conflict} find that children of internally displaced persons attain significantly lower levels of education; and FID has been linked to deteriorating mental health among both adults and adolescents \citep{tamayo2016problemas, leon2023mental, marroquin2020mental}.

Labor and housing markets are also affected. \citet{calderon2016labour} find that FID depresses the wages of unskilled urban workers who compete with forced migrants, although \citet{morales2018impact} shows that this effect tends to fade over time, except for low-skilled women. \citet{depetris2018unexpected} further show that FID inflows increase rental prices in low-income areas and reduce them in high-income neighborhoods.

Against this backdrop, a major shift occurred on December 20, 2014, when, after more than 60 years of internal conflict, the FARC declared a permanent unilateral ceasefire. This marked a turning point in the conflict: the FARC began withdrawing to more remote areas to avoid clashes with the Colombian army, setting the stage for the eventual peace agreement. This event presents a unique opportunity to analyze how the end of FARC’s direct involvement in hostilities affected FID.

In this paper, we investigate how the FARC’s unilateral ceasefire (and their subsequent retreat) impacted FID, comparing municipalities with prior FARC presence to those without. We leverage our identification strategy by exploiting the timing of the FARC’s announcement on December 20, 2014, together with the geographical distribution of FARC presence across municipalities prior to the ceasefire. The ceasefire is expected to lead to a decline in displacement. In particular, since FARC-related violence was primarily associated with large-scale displacement events \citep{cnmh2015}, we focus on the expected decrease in such events. Our results show that, after the FARC ceasefire, municipalities that had a FARC presence prior to the ceasefire were 17.0 percentage points less likely to exceed the upper quartile of the pre-ceasefire FID rate—a 22.8\% reduction from a pre-ceasefire mean of 74.5\%. Similarly, the probability of exceeding the upper decile of the pre-ceasefire FID rate fell by 20.9 percentage points, a significant 51.4\% decline from the pre-ceasefire mean of 40.7\%. These estimates indicate that the ceasefire had a meaningful impact on mitigating severe FID episodes.

Understanding the far-reaching consequences of the FARC’s ceasefire is essential for evaluating the broader impacts of peace processes on a wide range of socioeconomic and well-being indicators. Our paper contributes to the growing literature that examines the diverse effects of the FARC’s ceasefire across multiple domains. For example, \cite{prem2020end} find that areas controlled by the FARC prior to the declaration of a permanent ceasefire experienced a differential increase in deforestation after the ceasefire began. \cite{prem2022selective} show that the ceasefire also led to a surge in the targeting of community leaders in former FARC strongholds, perpetrated by armed groups excluded from the peace process seeking to consolidate control in those areas. In terms of education, \cite{prem2023human} find that areas most affected by FARC violence prior to the ceasefire experienced a large differential reduction in school dropout rates relative to other regions. Further analysis by \cite{bernal2024peaceful} reveals an 8 to 13 percent differential increase in the entry of new firms in municipalities formerly affected by FARC violence. In addition, \cite{guerra2024peace} observe a 3.2 percent increase in fertility in areas exposed to FARC-related violence compared to non-exposed areas. However, to the best of our knowledge, there is no existing evidence on the impact of the FARC’s unilateral ceasefire on FID.

The remainder of the paper is organized as follows. Section 2 provides the historical background. Section 3 presents the data sources, sample selection criteria, and definitions of the main variables, and provides summary statistics. Section 4 outlines the empirical strategy. Section 5 presents the main results, and Section 6 concludes.

\section{Historical Background} \label{Sec:hist}

The Revolutionary Armed Forces of Colombia (FARC) were founded in 1964 as the military wing of the Colombian Communist Party. Influenced by Marxist-Leninist ideology, the FARC sought to overthrow the Colombian government and redistribute wealth, particularly in rural areas plagued by severe economic inequality \citep{brittain2010revolutionary}. Throughout the 1970s and 1980s, the group expanded its operations, engaging in guerrilla warfare and securing financing through kidnapping, extortion, and, later, drug trafficking. Attempts at political participation, such as the creation of the Patriotic Union (UP) party in 1985, ultimately failed, after thousands of UP members were assassinated by paramilitaries and government-linked forces \citep{ramirez2011between}.

By the late 1990s and early 2000s, coinciding with the highest FID numbers, the FARC had reached its peak, controlling significant portions of Colombia's territory and posing a major threat to the state. In response, the government, with U.S. assistance through Plan Colombia\footnote{In July 2000, the United States approved a 1.3 billion USD aid package for Colombia, of which 80\% was allocated to strengthening the operational capacity of the Public Force (\url{https://www.comisiondelaverdad.co/el-plan-colombia}).}, intensified military operations against the group, weakening its command structure and reducing its territorial control \citep{tate2015drugs}. Under President Álvaro Uribe (2002–2010), counterinsurgency efforts resulted in the deaths of several top FARC leaders.

In 2012, during Juan Manuel Santos' first presidential term, the Colombian government and the FARC initiated peace negotiations. One key milestone during the peace negotiations was the unilateral permanent ceasefire declared by the FARC on December 20, 2014. While the FARC often declared temporary cessations of hostilities, especially during Christmas festivities, a permanent ceasefire was both unprecedented and unexpected. This move was intended to signal a clear commitment by the FARC to reach a peace agreement with the government \citep{guerra2024peace}. The FARC subsequently withdrew their troops to more remote areas where military contact with government security forces and other armed groups was unlikely to take place \citep{prem2022selective}. Although the FARC briefly suspended the ceasefire in May 2015, it was proclaimed again in July 2015, and the FARC’s offensive activities dropped by 98\% after December 2014, according to the Conflict Analysis Resource Center.\footnote{\url{https://blog.cerac.org.co/wp-content/uploads/2016/07/Reporte_MonitoreoDesescalamiento_Reporte12.pdf}} These patterns are consistent with the ceasefire being largely respected after the 2014 unilateral declaration, which was followed by a bilateral ceasefire agreement in August 2016 and the final peace agreement signed in November 2016 \citep{prem2022selective}. 

\section{Data} \label{Sec:data}
This section describes the data sources, sample selection criteria, and definitions of the main variables, and provides summary statistics.

\subsection{Data Sources}
To identify the FARC’s presence in the municipalities, we use data from SIVEL (Sistema de Información de Violencia Política y Derechos Humanos). SIVEL is integrated into the \href{https://cinep.org.co/}{Centro de Investigación y Educación Popular (CINEP)}, an NGO affiliated with the Society of Jesus, a religious order in the Catholic Church also known as the Jesuits. SIVEL tracks cases of forced disappearances, extrajudicial killings, and massacres. The main advantage of SIVEL is that it offers detailed descriptions of violent events, including the date of occurrence, the municipality in which the event took place, the identity of the perpetrator, and the number of victims involved \citep{restrepo2004dynamics}. Moreover, given the Catholic Church’s widespread presence across the country, even in remote areas—one can be confident in the coverage of these data \citep{restrepo2004dynamics}.

Data on FID come from the \href{https://www.unidadvictimas.gov.co/registro-unico-de-victimas-ruv/}{Registro Único de Víctimas (RUV)}. The Colombian government created the RUV under Law 1448 in 2011. The RUV serves as a central repository for documenting the experiences of victims, including those who have been forcibly displaced. By documenting detailed information on the causes and circumstances of displacement, the registry provides data that inform the planning of reparations, housing support, and legal assistance. We use the most recent data made publicly available by the RUV, which were last updated on December 31, 2024.

Finally, the Panel de Municipios, developed by \href{https://datoscede.uniandes.edu.co/catalogo-de-datos/}{CEDE (Centro de Estudios sobre Desarrollo Económico)} at the University of Los Andes in Colombia, provides information on municipal characteristics. It is a comprehensive, longitudinal, municipal-level database that compiles socioeconomic, demographic, and institutional data for all Colombian municipalities over time.

\subsection{Sample Selection and Main Indicators}

\begin{figure}[t!]
    \centering
        \caption{FARC presence between 2011-2014}
    \label{fig:map_farc}
    \includegraphics[width=0.5\linewidth]{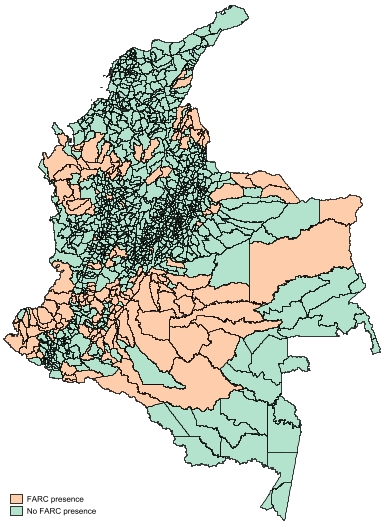}
        \begin{minipage}{0.5\linewidth} 
\centering \footnotesize Source: Authors' calculations using data from SIVEL.
\end{minipage}\\
\end{figure}

The FARC’s historical territorial focus was on rural and remote regions, particularly in areas with weak state presence, where it could secure financing through activities such as drug trafficking, extortion, and illegal mining. In contrast, the group maintained a low presence in departmental capitals with stronger institutional oversight \citep{ramirez2011between, international2017colombia}. This pattern is confirmed by \autoref{fig:map_farc}, which illustrates the geospatial distribution of the FARC’s presence across Colombian municipalities between 2011 and 2014 using the SIVEL database. Red-shaded areas indicate regions where the FARC was active, while green areas represent municipalities with no recorded presence. FARC’s activity was concentrated in rural and remote areas characterized by dense jungle terrain, coca cultivation, and historically weak state control. In contrast, municipalities in the central and northern regions, especially around the Andes and major urban centers, exhibit lower FARC activity, reflecting the group's rural-based insurgency strategy.

This geospatial distribution suggests that the effects of the ceasefire would manifest primarily in rural zones rather than urban centers. Hence, given that it is unlikely that FID originating in departmental capitals was a result of FARC presence, we do not consider departmental capitals in our analysis.

Our identification strategy exploits two key sources of variation: the temporal discontinuity created by the FARC's permanent ceasefire announcement on December 20, 2014, and the spatial variation in FARC presence across municipalities prior to this date. Given that the ceasefire was largely respected, we define the pre-ceasefire period as 2011-2014.

This temporal window is chosen to ensure policy consistency throughout the analysis period. President Juan Manuel Santos (2010-2018) maintained a diplomatic approach toward the FARC with the explicit goal of achieving a negotiated peace settlement, contrasting sharply with his predecessor Álvaro Uribe's militaristic strategy. We extend our post-ceasefire analysis through 2019 to capture Santos' complete presidency (and the initial year of President Iván Duque's term), ensuring that our results reflect the ceasefire's impact rather than confounding policy changes.

In order to identify municipalities that were exposed and those non-exposed to FARC activity before the ceasefire (that is, treated and control municipalities), we use SIVEL data. First, we determine the number of FARC-related events per 100,000 inhabitants in each municipality for each year during the 2011–2014 period. Next, we calculate the average number of events for the period for each municipality. Since the data indicate that both the median and the upper quartile are equal to 0, we classify municipalities as treated if they had at least one FARC-related violent event during the pre-ceasefire period.\footnote{This treatment definition is used, for example, in \cite{prem2020end}.} Therefore, control municipalities are those that had no violent event committed by the FARC during the same period.
 
Our outcome variables are defined as dummies that take the value of 1 if the FID rate (the number of FID cases per 100,000 inhabitants, hereafter "FID per 100k pop.") exceeds a given threshold and 0 otherwise. To empirically back up our threshold choices, it is worth looking at the "FID per 100k pop." distribution in our sample.

\begin{figure}[H]
    \centering
        \caption{FID per 100k pop. between 2011-2014}
    \label{fig:hist_fid}
    \includegraphics[width=0.75\linewidth]{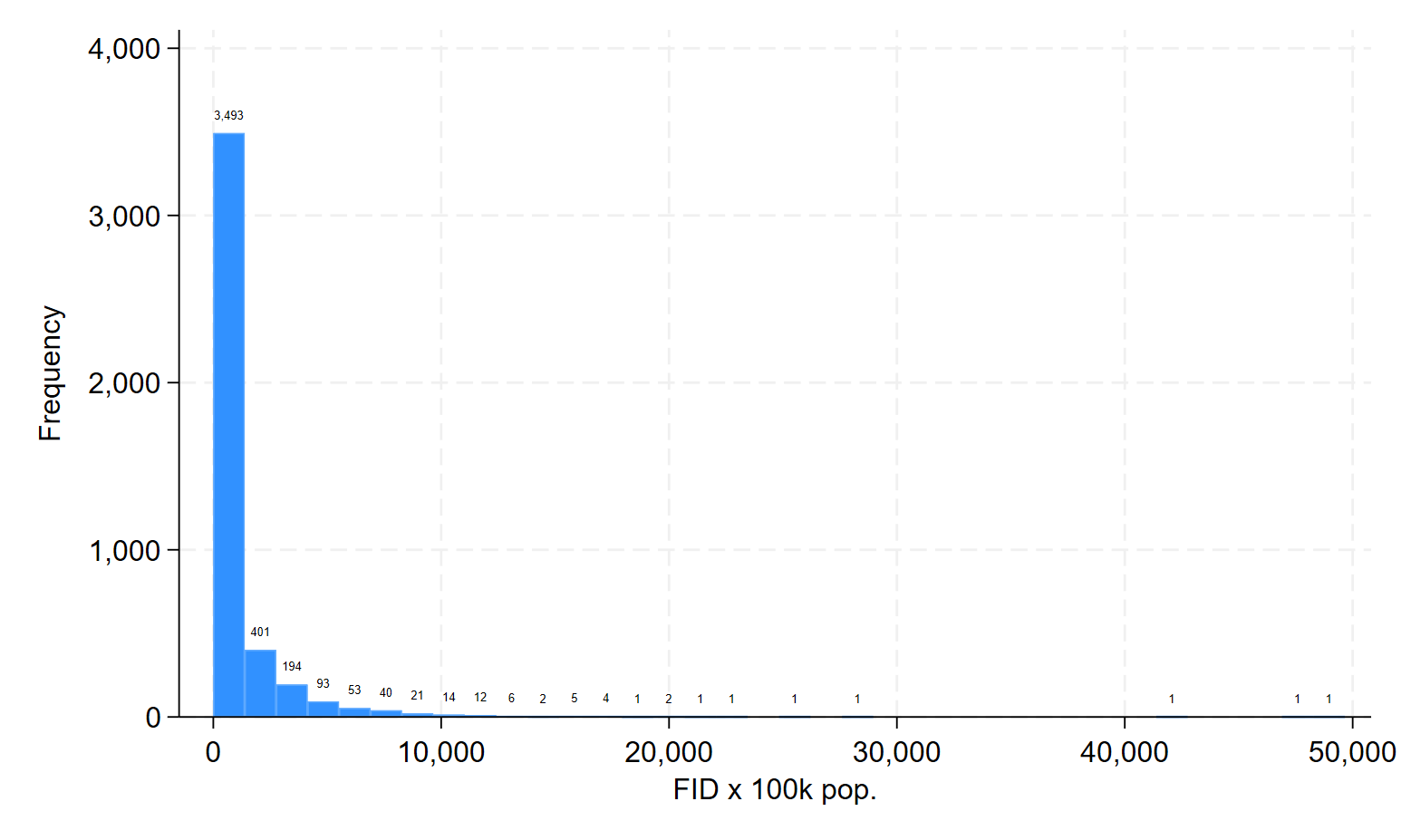}
        \begin{minipage}{0.75\linewidth} 
\centering \footnotesize Source: Authors' calculations using data from \textit{Registro Único de Victimas}.
\end{minipage}\\
\end{figure}

\autoref{fig:hist_fid} shows the "FID per 100k pop." histogram for the period 2011–2014. The histogram is heavily right-skewed, with a large number of observations clustered at a value of 0, indicating that many municipalities did not suffer from FID during the period analyzed. However, there is also a long right tail, indicating that some municipalities were highly affected by FID.

Since we are primarily interested in the effect of the ceasefire on reducing high displacement levels (which were more likely and frequently caused by the FARC), the thresholds chosen for our benchmark analyses are drawn from the upper tail of the FID rate distribution over all years in the pre-ceasefire period (2011–2014). Specifically, we use the 75th percentile (upper quartile) and the 90th percentile (upper decile). This allows us to assess whether the ceasefire reduced the probability of experiencing high displacement levels. Admittedly, given Colombia's persistent violence, displacement remains prevalent even in some municipalities without FARC presence. However, FARC-related violence is primarily associated with large-scale displacement events, making the upper tail of the distribution the most relevant for the analysis \citep{cnmh2015}.\footnote{In addition to this primary approach, we used an alternative way of defining the threshold using the FID rate distribution from the first year of the sample (2011). Both approaches yield similar results. Additionally, we estimated models using lower thresholds to define our outcomes, such as the median and the 33rd percentile, but the evidence suggests that the parallel trends assumption required for identification was not satisfied when using the corresponding outcomes.}

\subsection{Summary Statistics}

\autoref{tab:sum} provides summary statistics for all municipalities in our working sample, as well as for those with and without FARC activity during the pre-ceasefire period (2011–2014). Recall that, as previously explained, our analysis excludes departmental capitals, where FARC activity was significantly less relevant.
The total population, urban population, and rural population are all significantly higher in areas where FARC was active. Additionally, these areas are significantly larger in size, have a slightly higher percentage of rural population, and exhibit notably lower population density. FID and "FID per 100k pop." are substantially higher in municipalities affected by the FARC, consistent with FARC being a major driver of FID. These municipalities also have a higher probability of exceeding the upper quartile and upper decile of the FID distribution during the pre-ceasefire period. Specifically, 74\% and 41\% of municipalities with FARC activity surpassed the period's upper quartile and upper decile, respectively. This is consistent with the fact that these municipalities experienced events that led to large-scale displacement of residents from their home municipalities.

\begin{table}[htbp!]
\centering
\caption{Summary Statistics in the Pre-ceasefire Period (2011–2014)}
\label{tab:sum}
\def\sym#1{\ifmmode^{#1}\else\(^{#1}\)\fi}
\begin{tabularx}{\linewidth}{l *{4}{>{\centering\arraybackslash}X}}
\toprule
                & All       & No FARC  & FARC     & Diff.      \\
\midrule
Pop. Total      & 22,494.57 & 20,848.14 & 33,190.64 & \ensuremath{-12{,}342.49\sym{***}} \\
Pop. Urban      & 13,056.30 & 12,286.37 & 18,058.16 & \ensuremath{-5{,}771.78\sym{***}} \\
Pop. Rural      &  9,438.27 &  8,561.77 & 15,132.48 & \ensuremath{-6{,}570.71\sym{***}} \\
Area (Km.)      &    950.14 &    691.01 &  2,633.59 & \ensuremath{-1{,}942.58\sym{***}} \\
\% Rural Pop.   &      0.59 &      0.59 &      0.60 & \ensuremath{-0.01\sym{**}} \\
Pop. Density    &    117.80 &    127.11 &     57.29 & \ensuremath{69.83\sym{***}} \\
FID             &    243.93 &    118.94 &  1,055.93 & \ensuremath{-936.98\sym{***}} \\
FID x 100k pop. &  1,046.36 &    686.57 &  3,383.74 & \ensuremath{-2{,}697.17\sym{***}} \\
FID rate > upper quartile & 0.25 & 0.17 & 0.74 & \ensuremath{-0.57\sym{***}} \\
FID rate > upper decile   & 0.10 & 0.05 & 0.41 & \ensuremath{-0.35\sym{***}} \\
FARC            &     0.13 &     0.00 &     1.00 &             \\
\midrule
Observations    &    4348  &    3768  &     580  &     4348     \\
\bottomrule
\multicolumn{5}{p{\linewidth}}{\footnotesize 
* p < 0.10, ** p < 0.05, *** p < 0.01. "FARC" municipalities are defined as those that experienced at least one FARC-related violent episode during the pre-ceasefire period, while "No FARC" municipalities are those that did not. FID and "FID x 100k pop." (or FID rate) refer to the total number of FID cases and the total number of FID cases per 100,000 inhabitants in each municipality and year, respectively. "FID rate > upper quartile" and "FID rate > upper decile" are binary indicators equal to 1 if a municipality's yearly FID rate exceeds the upper quartile or the upper decile of the FID rate distribution across all years in the pre-ceasefire period (2011–2014), respectively.
} \\
\end{tabularx}
\end{table}

\section{Empirical Strategy} \label{Sec: empstrategy}

Following the existing literature analyzing the effects of the FARC ceasefire on other outcomes,\footnote{See, for example, \cite{prem2020end}, \cite{prem2022selective}, and \cite{prem2023human}.} we estimate the following difference-in-differences (DiD) model with fixed effects:

\begin{equation}
	\label{eq:main}
	y_{mt} = \alpha +\delta {FARC}_{m} \times Post_t + \varphi_m + \tau_{t} + \epsilon_{mt}
\end{equation}

where $y_{mt}$ is the outcome of interest, a dummy variable that takes the value 1 if the "FID per 100k pop." in municipality $m$ in year $t$ exceeds one of the previously discussed thresholds, and 0 otherwise. Note that $m$ represents the municipality from which individuals were displaced, not the host municipality. \footnote{While we have data on both the total number of individuals accommodated in and displaced from each municipality each year, we do not know the exact origin-and-destination municipality pairs, as this information is not publicly available.} $FARC_m$ is a dummy that takes the value 1 if municipality $m$ experienced at least one violent episode perpetrated by FARC between 2011 and 2014, and 0 otherwise. $Post_t$ is a dummy that equals 1 for years following the start of the permanent ceasefire (i.e., since 2015). Finally, $\varphi_m$ and $\tau_t$ are municipality and year fixed effects. Standard errors are clustered at the municipality level.

\begin{figure}[H]
    \centering    

    \caption{FID per 100k pop. above 2011–2014 Upper Quartile}
    \includegraphics[width=0.6\linewidth]{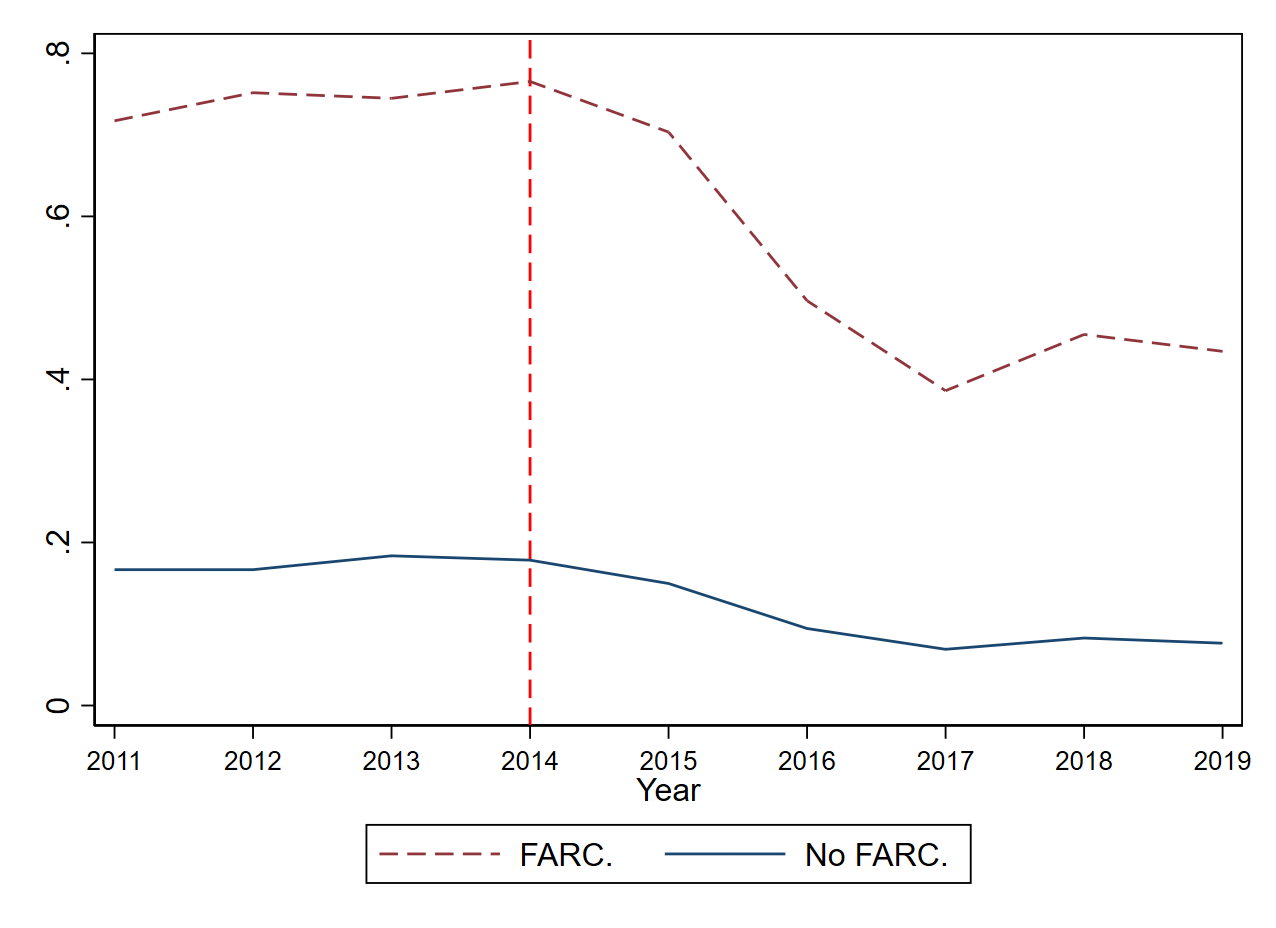}
    \label{fig:ts_p75}

    \caption{FID per 100k pop. above 2011–2014 Upper Decile}
    \includegraphics[width=0.6\linewidth]{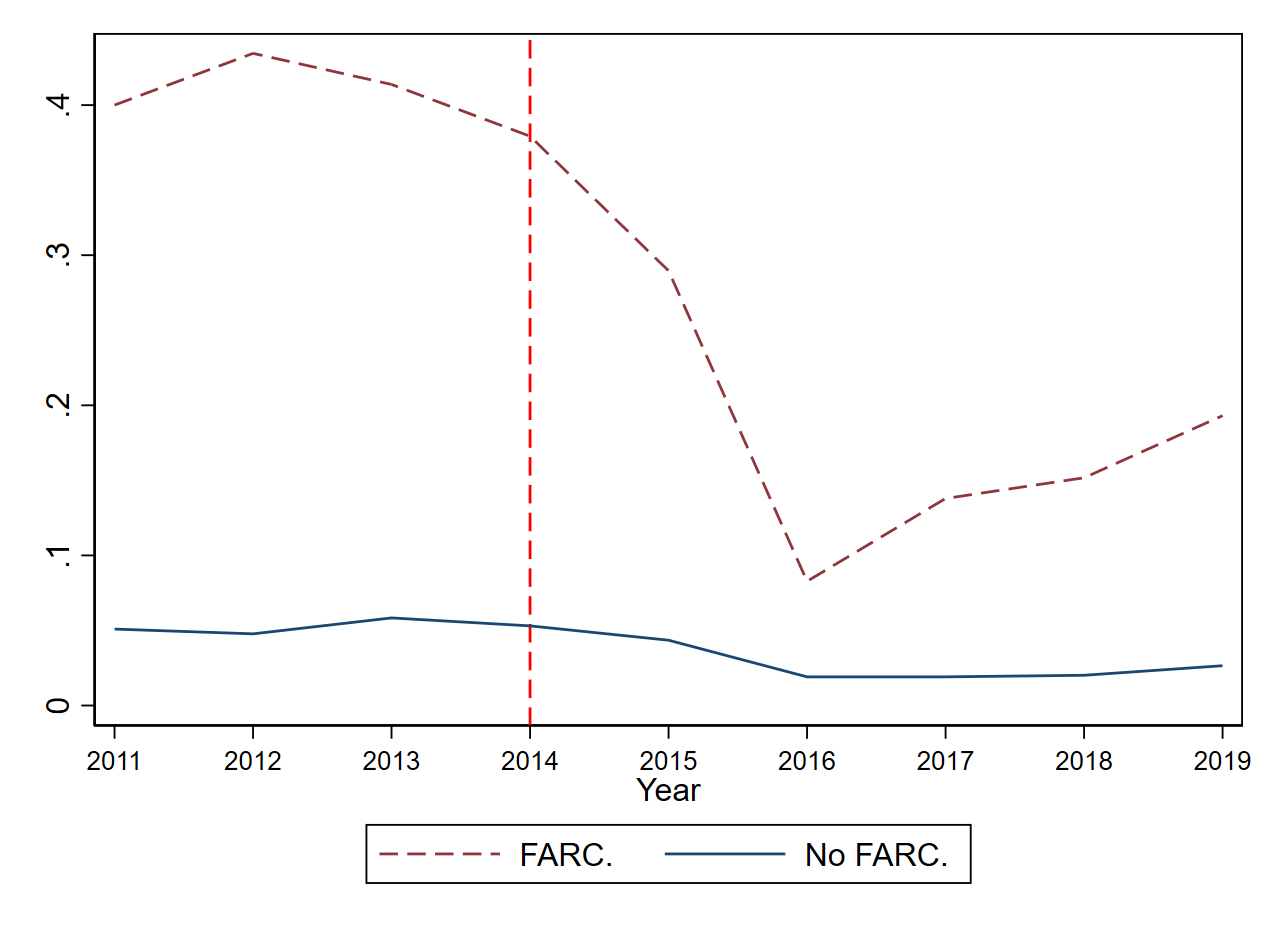}
    \label{fig:ts_p90}

    \caption*{
        \parbox{0.9\linewidth}{\footnotesize
        Note: The red dashed vertical line marks the year in which the FARC announced the unilateral permanent ceasefire (December 20, 2014). The figures show the yearly share of municipalities with an FID rate (number of FID cases per 100,000 inhabitants) that exceeded the upper quartile (\autoref{fig:ts_p75}) or the upper decile (\autoref{fig:ts_p90}) of the FID rate distribution across all years in the pre-ceasefire period. Shares are expressed on a 0–1 scale. “FARC” and “no FARC” refer to municipalities with and without FARC presence prior to the ceasefire.}
    }

\end{figure}

The key identifying assumption in a DiD framework is the parallel trends assumption. In our context, this means that, had there been no ceasefire, the FID outcomes would have followed parallel trajectories \textit{after} the ceasefire in municipalities with and without FARC activity prior to the ceasefire. While this counterfactual evolution is inherently unobservable, researchers typically assess the plausibility of the parallel trends assumption by examining pre-treatment trends. \autoref{fig:ts_p75} and \autoref{fig:ts_p90} present the two outcomes we consider. These figures indicate that municipalities with and without FARC activity exhibited similar outcome trajectories before the ceasefire, providing suggestive evidence in support of the validity of the parallel trends assumption. In the next section, we provide further suggestive evidence on the absence of pre-treatment differential trends via an event study analysis.\footnote{We also considered an extended version of \autoref{eq:main} that includes $Pop.Density_{mt}$ (total population over municipality size) and $RuralPop_{mt}$ (the percentage of the municipality population living in rural areas) as additional covariates. This specification relies on the conditional parallel trends assumption, which is weaker than its unconditional counterpart. However, since there is no evidence against the unconditional parallel trends assumption and the OLS estimator in covariate-augmented two-way fixed effects (TWFE) models may be biased \citep{Caetanoetal2022timevaryingCovariates, CaetanoCallaway2024conditionalPT}, we follow the recent literature and use a specification without additional covariates.}

It is also important to assess whether there are anticipation effects in DiD settings. In our study, individuals living in municipalities with FARC activity should not anticipate the permanent ceasefire declared by the FARC. \autoref{fig:ts_p75} and \autoref{fig:ts_p90} do not suggest any prominent changes in either definition of the outcome prior to the ceasefire. Therefore, it appears that the permanent ceasefire was not anticipated, consistent with the declaration being both unprecedented and unforeseen.

A key argument further supporting the (post-)ceasefire parallel trends assumption is the absence of major external factors that disproportionately affected one group of municipalities over the other after the ceasefire, in terms of FID outcomes. If no such differential shocks can be identified, this strengthens the credibility of the assumption that, in the absence of a ceasefire, municipalities with and without FARC activity would have continued to follow similar trajectories. Accordingly, the observed post-ceasefire differences can be more confidently attributed to the effect of the ceasefire itself. To the best of our knowledge, there were no other major social or political changes in Colombia around the time of the ceasefire declaration directly related to FID, such as large-scale military offensives, shifts in drug cartel dynamics, or significant expansions of paramilitary or criminal gang activity, that could have systematically influenced FID in a way that differentially affected municipalities with and without FARC activity.

\section{Results} \label{Sec:results}

\subsection{Main results}

\autoref{tab:main} shows the main results. Our coefficient of interest is $\delta$ in \autoref{eq:main}, the coefficient on the interaction between the binary indicator identifying municipalities with FARC activity during the pre-ceasefire period (2011–2014), $FARC_m$, and another binary indicator, $Post_t$, which takes the value 1 during the period following the announcement of the FARC's unilateral ceasefire. We present results for two outcomes that reflect high displacement levels: “FID rate > upper quartile” (Column 1) and “FID rate > upper decile” (Column 2), which are binary indicators equal to 1 if a municipality's yearly FID rate exceeds the upper quartile and the upper decile of the FID rate distribution across all years in the pre-ceasefire period, respectively. This allows us to examine whether, and to what extent, the ceasefire affected municipalities' probability of experiencing high levels of FID.

The results show that the FARC's ceasefire significantly reduced the probability of experiencing high displacement levels. Column (1) indicates that exceeding the upper FID quartile rate became 17 percentage points less likely after 2014, a 22.8\% relative reduction from the pre-ceasefire mean, when 74.4\% of the municipalities with FARC presence had an FID rate above that threshold. In Column (2), we show that the probability of exceeding the upper FID decile decreased by 20.9 percentage points, a sharp 51.4\% reduction from the pre-ceasefire mean of 40.7\%. The magnitude of these effects suggests that the ceasefire had a substantial impact on preventing high FID levels.

\begin{table}[H]\centering
\def\sym#1{\ifmmode^{#1}\else\(^{#1}\)\fi}
\caption{The Impact of the FARC's Ceasefire on FID}
\label{tab:main}
\begin{tabularx}{\linewidth}{l*{2}{>{\centering\arraybackslash}X}}
\toprule
                    & (1) & (2) \\
                    & FID rate > upper quartile & FID rate > upper decile \\
\midrule
FARC $\times$ Post  & -0.170\sym{***} & -0.209\sym{***} \\
                    & (0.029)         & (0.032)         \\
\hline
Municipality F.E.    & Yes             & Yes             \\
Year F.E.           & Yes             & Yes             \\
\hline
Mean Dep. Var.  & 0.745 & 0.407 \\
(FARC municipalities, pre-ceasefire) &  &  \\
S.D. Dep. Var.  & 0.436 & 0.492 \\
(FARC municipalities, pre-ceasefire) & &  \\
\hline
Municipalities      & 1,087 & 1,087 \\
Adjusted R$^2$      & 0.69  & 0.52  \\
Obs.                & 9,783 & 9,783 \\
\bottomrule
\multicolumn{3}{p{\linewidth}}{\footnotesize Note: This table shows OLS coefficient estimates of $\delta$ from \autoref{eq:main}. FARC municipalities are defined as those that experienced at least one FARC-related violent episode during the pre-ceasefire period, and $Post$ takes the value 1 during the period following the announcement of the FARC's unilateral ceasefire. "FID rate > upper quartile" and "FID rate > upper decile" are binary indicators equal to 1 if a municipality's yearly FID rate (total number of FID cases per 100,000 inhabitants) exceeds the upper quartile or the upper decile of the FID rate distribution across all years in the pre-ceasefire period, respectively. Clustered standard errors at the municipality level are in parentheses. * p < 0.10, ** p < 0.05, *** p < 0.01.}\\
\end{tabularx}
\end{table}

\subsection{Event-study Analysis}

We now extend our investigation to perform an event-study analysis. In practice, the event study in our case consists of estimating a dynamic version of the DiD model, where the effect of the ceasefire is captured by a series of lead and lag coefficients corresponding to different time periods relative to the ceasefire. We estimate the following equation:

\begin{equation}
\label{eq:pt}
y_{mt} = \sum_{t \neq 2014 }\rho_t (d_t \times \text{{FARC}}_{m}) + \varphi_{m} + \gamma_{t} + \varepsilon_{mt}
\end{equation}

where $y_{mt}$ denotes our outcomes of interest, i.e., binary variables that take the value 1 if the number of FID cases per 100,000 inhabitants in municipality $m$ in year $t$ exceeds the upper quartile or the upper decile of the FID rate distribution across all years in the pre-ceasefire period (2011–2014), respectively. $FARC_m$ is a dummy that takes the value 1 if municipality $m$ experienced at least one violent episode perpetrated by the FARC between 2011 and 2014, and 0 otherwise. $d_t$ is a dummy variable that equals 1 for each year $t$ of the sample period, $\varphi_m$ are municipality fixed effects, and $\gamma_{t}$ are year fixed effects. $\varepsilon_{mt}$ is the error term. Standard errors are clustered at the municipality level.

The event-study analysis serves two main purposes. First, it allows for a more flexible evaluation of the effect of the ceasefire over time \citep{roth2023s} by enabling the estimation of dynamic effects. That is, instead of assuming a constant (average) effect, an event study captures the evolution of the impact of the ceasefire over time. This allows us to assess whether the impact persisted, increased, or faded out after the ceasefire. Second, an event study is also useful for providing further suggestive evidence on the validity of the DiD identifying assumptions. In particular, examining the effects in pre-ceasefire periods is useful for detecting anticipation effects (i.e., changes in FID before the intervention). Additionally, by examining the estimated effects for pre-ceasefire periods, we can assess whether our FID outcomes followed parallel trends across municipalities with and without FARC activity prior to the ceasefire. If pre-ceasefire impacts are close to zero and statistically insignificant, this is viewed as further suggestive evidence that post-ceasefire FID would have followed parallel paths over time had there been no ceasefire, thus supporting the parallel trends assumption.

\begin{figure}[b!]
    \centering    
  \caption{The Impact of the FARC’s Ceasefire on FID-Upper Quartile. Event-study Results}
    \includegraphics[width=0.6\linewidth]{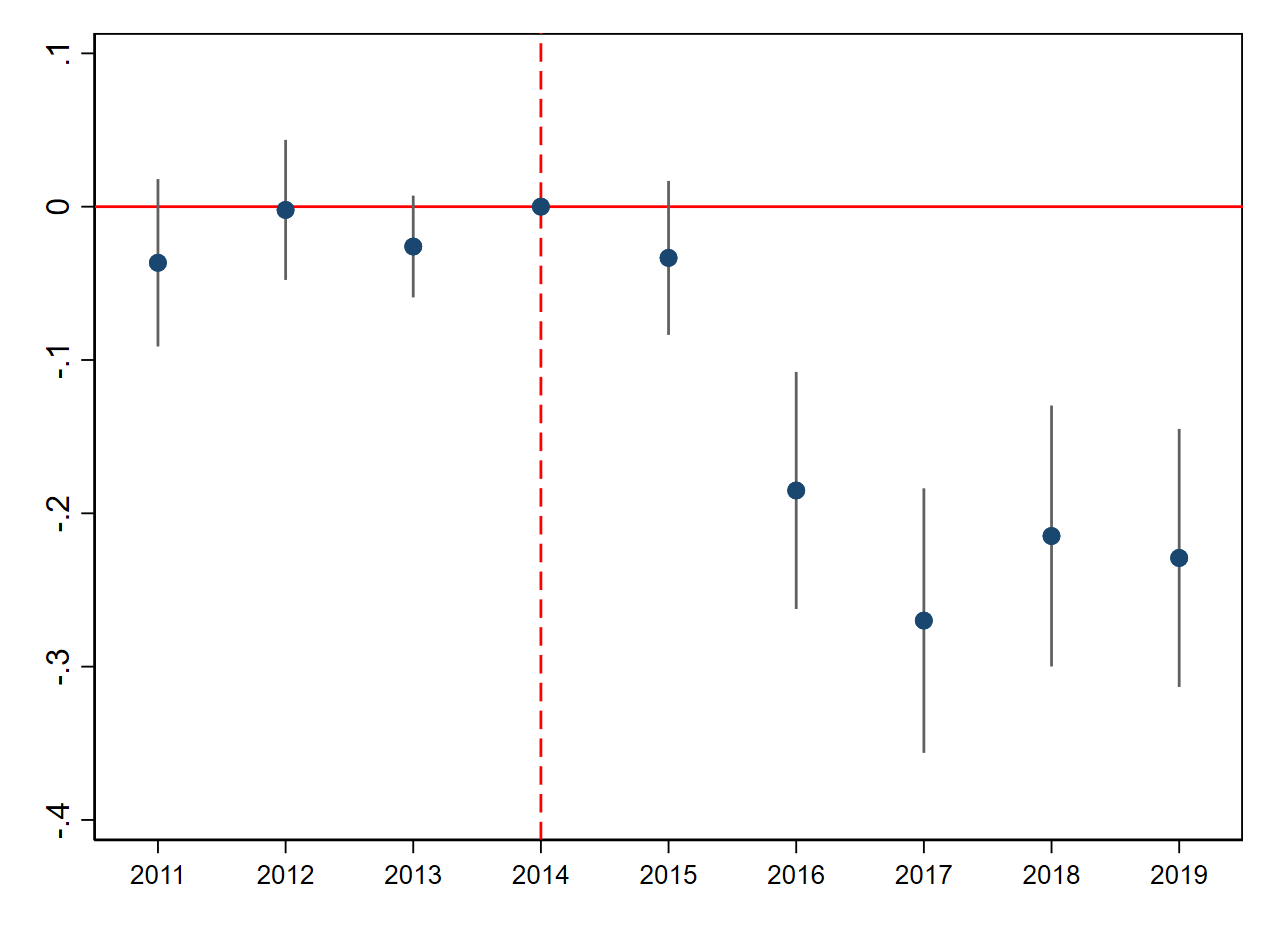}
    \label{fig:pt_p75}
      \caption{The Impact of the FARC’s Ceasefire on FID-Upper Decile. Event-study Results}
    \includegraphics[width=0.6\linewidth]{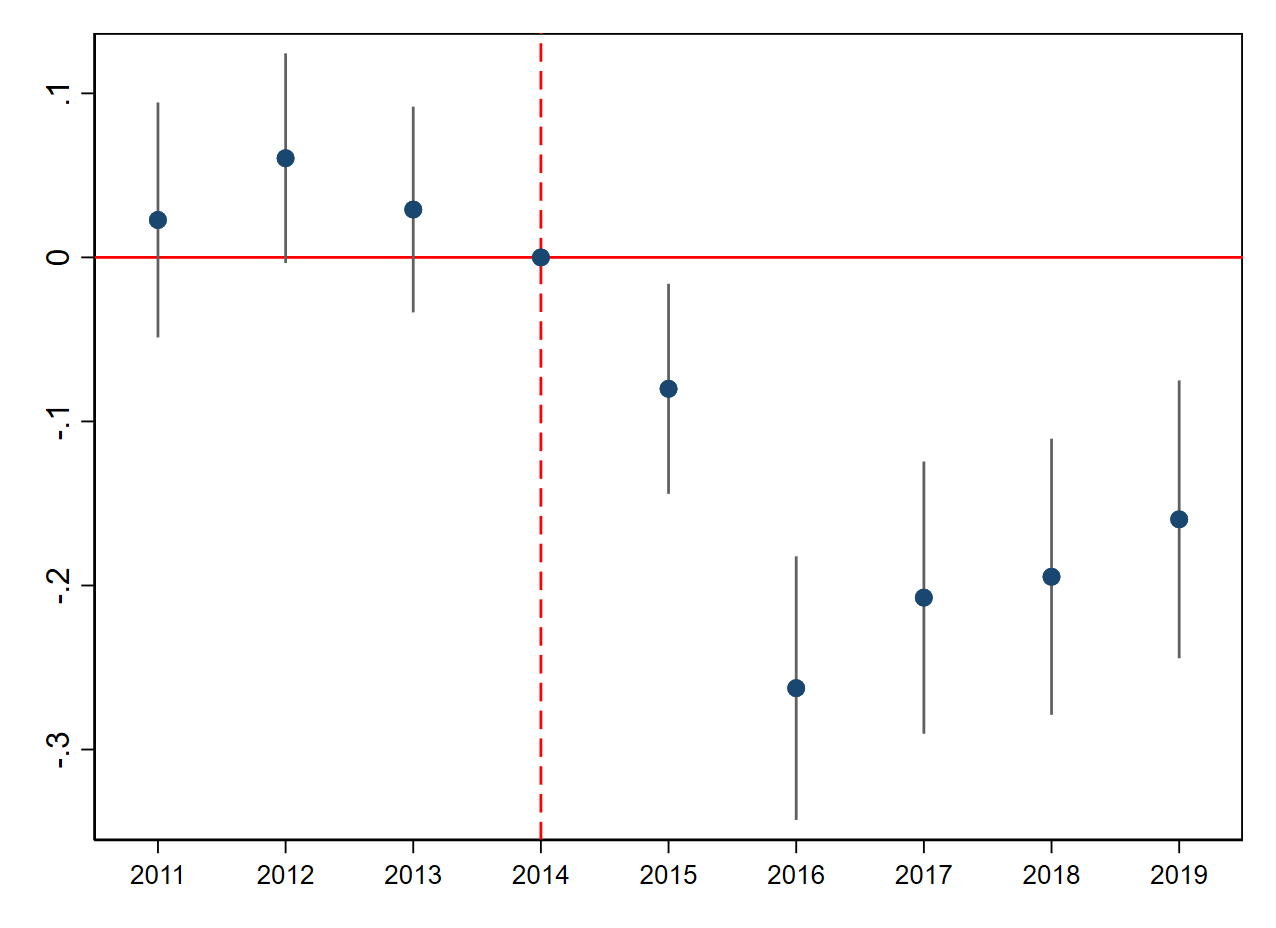}
    \label{fig:pt_p90}

     \caption*{
        \parbox{0.9\linewidth}{\footnotesize
        Note: The red dashed vertical line marks the year in which the FARC announced the unilateral permanent ceasefire (December 20, 2014). The figures plot the event-study coefficient estimates ($\hat{\rho}_t$) from \autoref{eq:pt} along with their corresponding 95\% confidence intervals for our two outcomes of interest: binary indicators equal to 1 if a municipality's yearly FID rate (number of FID cases per 100,000 inhabitants) exceeds the upper quartile (\autoref{fig:pt_p75}) or the upper decile (\autoref{fig:pt_p90}) of the FID rate distribution across all years in the pre-ceasefire period.} }

\end{figure}

\autoref{fig:pt_p75} and \autoref{fig:pt_p90} plot the event-study coefficient estimates ($\hat{\rho}_t$) from \autoref{eq:pt} along with their corresponding 95\% confidence intervals for our two outcomes of interest. \autoref{tab:pt_fid_mean114} in the appendix contains the full set of estimation results. Reassuringly, the pre-ceasefire coefficient estimates are not statistically significant for both outcomes, suggesting that, consistent with the unexpected nature of the ceasefire declaration, there are no anticipation effects. This finding also supports the plausibility of the parallel trends assumption and is typically interpreted as suggestive evidence in favor of the validity of the DiD design \citep{roth2023s}.

The estimated coefficients for the post-ceasefire period reveal a gradual onset of its effect on FID, which then becomes larger and remains persistent in the medium and long term. In the first year after the ceasefire, the estimated effects display a negative sign, as expected, but they are relatively small and statistically insignificant in \autoref{fig:pt_p75}. This suggests that the ceasefire did not lead to an immediate and substantial reduction in FID in municipalities with FARC presence relative to their no-FARC activity counterparts. This is consistent with the uncertainty surrounding the ceasefire in 2015, when, despite a 98\% drop in FARC offensive activity compared to 2014, the truce was briefly suspended between May and June, which could have prompted individuals to flee exposed municipalities to a degree similar to pre-ceasefire levels. From 2016 onward, however, the estimated reductions in FID became markedly larger and statistically significant. In summary, the estimated coefficients confirm a somewhat delayed but sizable and persistent effect of the ceasefire on FID.

\subsection{Placebo tests}

Following \cite{bernal2024peaceful} and \cite{guerra2024peace}, we conduct a series of placebo exercises by estimating the main specification (\autoref{eq:main}) restricted to the pre-ceasefire period (2011–2014) and artificially assigning the ceasefire date to years in which it did not actually occur. This falsification test allows us to assess whether the estimated effects are driven by mere chance, pre-existing trends, or unaccounted factors rather than the actual ceasefire.

To perform these tests, we define three placebo dummy variables that take the value of 1 in 2012, 2013, and 2014, and estimate \autoref{eq:main} using only data from the actual pre-ceasefire period, substituting $Post$ (which identifies the true post-ceasefire periods) with the three placebo dummies. The results of these estimations, presented in \autoref{tab:placebo2012}, \autoref{tab:placebo2013}, and \autoref{tab:placebo2014}, are reassuring because the estimated placebo effects are both small and far from reaching standard levels of statistical significance. Hence, this evidence suggests that our results are genuine rather than driven by type I error or spurious factors.

\section{Conclusion} \label{conclusion}

Internal armed conflicts, particularly those involving multiple actors, often lead to large-scale forced internal displacement (FID). Colombia’s prolonged conflict has made the country one of the most affected by FID in the world.

This study examines the impact of the unilateral and permanent ceasefire declared by the FARC in 2014 and their subsequent withdrawal on FID. To investigate this, we use a difference-in-differences identification strategy that exploits the timing of the FARC’s announcement in 2014, together with the geographical distribution of FARC presence across municipalities prior to the ceasefire. In a nutshell, we find that the ceasefire and withdrawal had a substantial and significant impact on the reduction of severe FID episodes.

More specifically, we find that after the FARC ceasefire, municipalities that had a FARC presence prior to the ceasefire were 17 percentage points less likely to exceed the upper quartile of the pre-ceasefire FID rate, representing a 22.8\% reduction relative to the pre-ceasefire mean of 74.5\%. In the same vein, the probability of exceeding the upper decile of the pre-ceasefire FID rate declined by 20.9 percentage points, a significant 51.4\% reduction relative to the pre-ceasefire mean of 40.7\%. Additionally, our event-study analysis indicates that the ceasefire’s impact on FID emerged gradually, with large and persistent effects in the medium to long term.

These results highlight the importance of stability and the effective implementation of peace agreements in mitigating forced displacement and its consequences. Understanding the determinants of FID is crucial due to its profound humanitarian, economic, and social implications, which affect not only displaced individuals across many dimensions\footnote{See, for instance, \cite{ibanez2008civil}, \cite{wharton2011conflict}, \cite{tamayo2016problemas}, \cite{leon2023mental}, \cite{marroquin2020mental}, \cite{calderon2016labour}, and \cite{depetris2018unexpected}.} but also receiving populations and host communities \citep{becker2019consequences, verme2021impact}.

From a policy perspective, our results point to the importance of strengthening state presence in historically conflict-affected territories and underscore the need for public policies that ensure a sustainable transition to peace while mitigating the adverse effects of displacement. This involves not only ensuring security in areas formerly controlled by armed groups, but also investing in infrastructure, education, and public services to facilitate the integration of displaced populations and the reconstruction of social cohesion. Additionally, monitoring the power vacuum left by the FARC’s withdrawal is essential, as it could encourage the expansion of other armed actors and perpetuate cycles of violence and displacement \citep{prem2022selective}.

\bibliography{Bibtes}

\part*{Appendices}

\setcounter{equation}{0}\renewcommand{\theequation}{\Alph{section}.\arabic{equation}}
\setcounter{figure}{0}\renewcommand{\thefigure}{\Alph{section}.\arabic{figure}}
\setcounter{table}{0}\renewcommand{\thetable}{\Alph{section}.\arabic{table}}

\begin{appendices}
\renewcommand\thesection{Appendix \Alph{section}}
\crefalias{section}{appsec}

\section{Additional Tables}\label{app:hhchar}

\begin{table}[htbp!]
\centering
\def\sym#1{\ifmmode^{#1}\else\(^{#1}\)\fi}
\caption{The Impact of the FARC’s Ceasefire on FID. Event-study Estimates}
\label{tab:pt_fid_mean114}
\begin{tabularx}{\linewidth}{p{8cm}*{2}{X}}
\toprule
                   & (1) & (2) \\
                    & FID rate > upper quartile & FID rate > upper decile \\
        
\midrule
Year=2011 $\times$ FARC&      -0.037         &       0.023         \\
                    &     (0.028)         &     (0.036)         \\
Year=2012 $\times$ FARC&      -0.002         &       0.060\sym{*}  \\
                    &     (0.023)         &     (0.033)         \\
Year=2013 $\times$ FARC&      -0.026         &       0.029         \\
                    &     (0.017)         &     (0.032)         \\
Year=2015 $\times$ FARC&      -0.033         &      -0.080\sym{**} \\
                    &     (0.026)         &     (0.033)         \\
Year=2016 $\times$ FARC&      -0.185\sym{***}&      -0.263\sym{***}\\
                    &     (0.039)         &     (0.041)         \\
Year=2017 $\times$ FARC&      -0.270\sym{***}&      -0.207\sym{***}\\
                    &     (0.044)         &     (0.042)         \\
Year=2018 $\times$ FARC&      -0.215\sym{***}&      -0.195\sym{***}\\
                    &     (0.043)         &     (0.043)         \\
Year=2019 $\times$ FARC&      -0.229\sym{***}&      -0.160\sym{***}\\
                    &     (0.043)         &     (0.043)         \\
                    \midrule
Municipality F.E.   &       Yes             &   Yes \\
Year F.E. &       Yes             &   Yes \\
\midrule
Mean Dep. Var. &       0.745         &       0.407         \\
(FARC municipalities, pre-ceasefire) &  &  \\
S.D. Dep. Var. &       0.436         &       0.492         \\
(FARC municipalities, pre-ceasefire) &  &  \\
\midrule
Municipalities      &       1,087         &       1,087         \\
Adjusted R$^2$      &        0.69         &        0.53         \\
Obs.                &       9,783         &       9,783         \\

\bottomrule
\multicolumn{3}{p{\linewidth}}{\footnotesize Note: This table shows OLS event-study coefficient estimates ($\hat{\rho}_t$) from \autoref{eq:pt} for our two outcomes of interest: binary indicators equal to 1 if a municipality's yearly FID rate (number of FID cases per 100,000 inhabitants) is above the upper quartile (column 1) or the upper decile (column 2) of the FID rate distribution across all years in the pre-ceasefire period. FARC municipalities are defined as those that recorded at least one FARC-related violent episode during the pre-ceasefire period. Clustered standard errors at the municipality level appear in parentheses. * p < 0.10, ** p < 0.05, *** p < 0.01.}\\
\end{tabularx}
\end{table}

\begin{table}[htbp]\centering
\def\sym#1{\ifmmode^{#1}\else\(^{#1}\)\fi}
\caption{Placebo Test: Ceasefire after 2011}
\label{tab:placebo2012}
\begin{tabularx}{\linewidth}{l*{2}{>{\centering\arraybackslash}X}}
\toprule
                    & (1) & (2) \\
                    & FID rate > upper quartile & FID rate > upper decile \\
\midrule
FARC $\times$ Placebo 2012&       0.026         &       0.015         \\
                    &     (0.022)         &     (0.028)         \\
\hline
Municipality F.E.   & Yes             & Yes             \\
Year F.E.           & Yes             & Yes             \\
\hline
Mean Dep. Var.  & 0.738 & 0.400 \\
(FARC municipalities, pre-2012) &   &  \\
S.D. Dep. Var.  & 0.441 & 0.492 \\
(FARC municipalities, pre-2012) & &  \\
\hline
Municipalities      &       1,087         &       1,087         \\
Adjusted R$^2$      &        0.83         &        0.72         \\
Obs.                &       4,348         &       4,348         \\

\bottomrule
\multicolumn{3}{p{\linewidth}}{\footnotesize Note: This table shows OLS coefficient estimates of $\delta$ from \autoref{eq:main}, restricting the sample to the pre-ceasefire period (2011–2014) and substituting $Post$ with “Placebo 2012,” which takes the value of 1 after 2011. FARC municipalities are defined as those that experienced at least one FARC-related violent episode during the pre-ceasefire period. “FID rate > upper quartile” and “FID rate > upper decile” are binary indicators equal to 1 if a municipality’s yearly FID rate (total number of FID cases per 100,000 inhabitants) exceeds the upper quartile or the upper decile of the FID rate distribution across all years in the pre-ceasefire period, respectively. Clustered standard errors at the municipality level are in parentheses. * p < 0.10, ** p < 0.05, *** p < 0.01.}\\
\end{tabularx}
\end{table}

\begin{table}[htbp]\centering
\def\sym#1{\ifmmode^{#1}\else\(^{#1}\)\fi}
\caption{Placebo Test: Ceasefire after 2012}
\label{tab:placebo2013}
\begin{tabularx}{\linewidth}{l*{2}{>{\centering\arraybackslash}X}}
\toprule
                    & (1) & (2) \\
                    & FID rate > upper quartile & FID rate > upper decile \\
\midrule
FARC $\times$ Placebo 2013&       0.018         &      -0.034           \\
                    &     (0.024)         &     (0.028)         \\
\hline
Municipality F.E.    & Yes             & Yes             \\
Year F.E.           & Yes             & Yes             \\
\hline
Mean Dep. Var.  &       0.745         &       0.424         \\
(FARC municipalities, pre-2013) &   &  \\
S.D. Dep. Var.  &       0.437         &       0.495         \\
(FARC municipalities, pre-2013) & &  \\
\hline
Municipalities      &       1,087         &       1,087         \\
Adjusted R$^2$      &        0.83         &        0.72         \\
Obs.                &       4,348         &       4,348         \\

\bottomrule
\multicolumn{3}{p{\linewidth}}{\footnotesize Note: This table shows OLS coefficient estimates of $\delta$ from \autoref{eq:main}, restricting the sample to the pre-ceasefire period (2011–2014) and substituting $Post$ with “Placebo 2013,” which takes the value of 1 after 2012. FARC municipalities are defined as those that experienced at least one FARC-related violent episode during the pre-ceasefire period. “FID rate > upper quartile” and “FID rate > upper decile” are binary indicators equal to 1 if a municipality’s yearly FID rate (total number of FID cases per 100,000 inhabitants) exceeds the upper quartile or the upper decile of the FID rate distribution across all years in the pre-ceasefire period, respectively. Clustered standard errors at the municipality level are in parentheses. * p < 0.10, ** p < 0.05, *** p < 0.01.}\\
\end{tabularx}
\end{table}

\begin{table}[htbp]\centering
\def\sym#1{\ifmmode^{#1}\else\(^{#1}\)\fi}
\caption{Placebo Test: Ceasefire after 2014}
\label{tab:placebo2014}
\begin{tabularx}{\linewidth}{l*{2}{>{\centering\arraybackslash}X}}
\toprule
                    & (1) & (2) \\
                    & FID rate > upper quartile & FID rate > upper decile \\
\midrule
FARC $\times$ Placebo 2014&       0.015         &      -0.039           \\
                    &     (0.017)         &     (0.028)         \\
\hline
Municipality F.E.    & Yes             & Yes             \\
Year F.E.           & Yes             & Yes             \\
\hline
Mean Dep. Var.  &       0.745         &       0.416         \\
(FARC municipalities, pre-2014) &   &  \\
S.D. Dep. Var.  &       0.436         &       0.493         \\
(FARC municipalities, pre-2014) & &  \\
\hline
Municipalities      &       1,087         &       1,087         \\
Adjusted R$^2$      &        0.84         &        0.72         \\
Obs.                &       4,348         &       4,348         \\

\bottomrule
\multicolumn{3}{p{\linewidth}}{\footnotesize Note: This table shows OLS coefficient estimates of $\delta$ from \autoref{eq:main}, restricting the sample to the pre-ceasefire period (2011–2014) and substituting $Post$ with “Placebo 2014,” which takes the value of 1 after 2013. FARC municipalities are defined as those that experienced at least one FARC-related violent episode during the pre-ceasefire period. “FID rate > upper quartile” and “FID rate > upper decile” are binary indicators equal to 1 if a municipality’s yearly FID rate (total number of FID cases per 100,000 inhabitants) exceeds the upper quartile or the upper decile of the FID rate distribution across all years in the pre-ceasefire period, respectively. Clustered standard errors at the municipality level are in parentheses. * p < 0.10, ** p < 0.05, *** p < 0.01.}\\
\end{tabularx}
\end{table}

\end{appendices}

\end{document}